\documentclass[twocolumn,prl,showpacs,preprintnumbers,amsmath,amssymb]{revtex4}
\usepackage{graphicx}% Include figure files
\usepackage{dcolumn}% Align table columns on decimal point
\usepackage{bm}% bold math

\begin{document}
\author{Jian Qi Shen \footnote{E-mail address: jqshen@coer.zju.edu.cn}}
\address{Zhejiang Institute of Modern Physics and Department
of Physics, Zhejiang University (Yuquan Campus), Hangzhou 310027,
People's Republic of China}
\date{\today}
\title{Nonanalytic metric in the presence of gravitomagnetic monopoles}

\begin{abstract}
The {\it nonanalytic} property of metric resulting from the
presence of gravitomagnetic monopoles is considered. The curvature
tensors, dual curvature tensors, dual Einstein tensor (and hence
the gravitational field equation of gravitomagnetic matter)
expressed in terms of {\it nonanalytic} metric are analyzed. It is
shown that the {\it spinor gravitomagnetic monopole} may be one of
the potential origins of the cosmological constant. An alternative
approach to the cosmological constant problem is thus proposed
based on the concept of gravitomagnetic monopole.
\end{abstract}
\maketitle

 The gravitomagnetic interaction attracts
attention of researchers in various areas such as the neutron
interferometry experiment \cite{Overhauser,Werner,Anandan},
gravitational geometric phase shift
\cite{Dresden,Furtado,Wagh,Shenprb}, interaction between moving
particles and gravitomagnetic fields \cite{Hehl,Ahmedov} as well
as spin-rotation coupling
\cite{Mashhoon1,Mashhoon2,Mashhoon3,Shenscript} (or the coupling
of gravitomagnetic moment to gravitomagnetic fields \cite{Shen3}).
Basically, the interactions involved in these considerations are
all in connection with the contribution of gravitomagnetic dipole
moments. As for the gravitomagnetic monopole, it has been
investigated by many authors in the literature since 1960's. These
studies include the spherically and cylindrically symmetric metric
produced by gravitomagnetic charge
 \cite{Newman,Demiansky,Nouri-Zonoz}, various properties and
effects ({\it e.g.}, geodesics, trajectory, and gravitational
lensing effect) in NUT spacetime
 \cite{Zimmerman,Lynden-Bell,Miller,Bini2002}, the differential geometric
and quantal field properties  \cite{Bini1,Bini2} as well as the
quantization of gravitomagnetic charge \cite{Dowker,Dowker2,Zee}.
Gravitomagnetic charge is a kind of topological (dual) charge,
which can be referred to as ``{\it dual mass}'' \cite{Shen}. Some
authors referred to it as ``{\it magnetic mass}'' or {\it
magnetic-type mass} ({\it magnetic-like mass}) \cite{Zimmerman}.
In this sense, matter may be classified into two categories, {\it
i.e.}, the regular matter (gravitoelectric matter) and the dual
matter (gravitomagnetic matter). The former has {\it mass} while
the latter has {\it dual mass}. More recently, we considered some
related topics of such a topological dual mass, including the
geometric (topological) effects in NUT space, dual curvature
tensors, dual current density as well as gravitational field
equation of gravitomagnetic monopole of its own \cite{Shen,Shen2}.

As the gravitomagnetic charge (should such exist) is a kind of
topological charge, which leads to some new interesting geometric
(topological) effects of spacetime, one should first examine in
detail the topological (global) properties of metric, where the
gravitomagnetic matter is present. It is found that if the
gravitomagnetic charge exists in spacetime, then the metric is no
longer analytic. One of the most immediate consequences due to
this property is such that the partial derivatives of the metric
with respect to the spacetime coordinates cannot commute, {\it
i.e.}, $\partial_{\alpha}\partial_{\beta}g_{\mu\nu}\neq
\partial_{\beta}\partial_{\alpha}g_{\mu\nu}$.
Although many properties relevant to the metric of gravitomagnetic
monopole have been considered in the literature
\cite{Zimmerman,Lynden-Bell,Miller}, to the best of our knowledge,
the {\it nonanalytic} property of metric in the presence of
gravitomagnetic monopoles may have so far never been discussed in
detail. In the present Letter, we further consider two subjects:
i) the curvature tensors and dual curvature tensors with {\it
nonanalytic} metric; ii) the connection between matter and dual
matter in gravitational field equations. As a result of this
consideration, we show that the {\it spinor gravitomagnetic
matter} may be one of the potential origins of the cosmological
constant. Thus the study of the dynamics of gravitomagnetic
monopoles might open up a possibility of probing into the
cosmological constant problem \cite{Weinberg2}.

In order to study the nonanalytic metric as well as the dynamics
of gravitomagnetic monopole, we should first define a dual
Riemannian curvature tensor as follows \cite{Shen2}
\begin{equation}
\widetilde{{\mathcal
R}}_{\mu\tau\omega\nu}=\frac{1}{2}\epsilon_{\mu\tau}^{\ \ \
\lambda\sigma}{\mathcal R}_{\lambda\sigma\omega\nu},
\label{dualRiemann}
\end{equation}
where $\epsilon_{\mu\tau}^{\ \ \ \lambda\sigma}$ denotes the
Levi-Civita tensor. Note that such a dual curvature tensor is
antisymmetric in both indices ($\mu, \tau$) and ($\omega, \nu$).
It follows from the definition (\ref{dualRiemann}) that there
exists a duality relationship between the Riemannian curvature
tensor and its dual, {\it i.e.},
\begin{equation}
{\mathcal R}_{\mu\tau\omega\nu}=-\frac{1}{2}\epsilon_{\mu\tau}^{\
\ \ \lambda\sigma}\widetilde{{\mathcal
R}}_{\lambda\sigma\omega\nu}. \label{dualRiemann2}
\end{equation}
Thus one can obtain a connection between the curvature scalar
${\mathcal R}$ and the dual curvature scalar $\widetilde{{\mathcal
R}}$. This connection is expressed as follows
\begin{equation}
{\mathcal
R}=-\frac{1}{2}\epsilon^{\nu\mu\tau\omega}\widetilde{{\mathcal
R}}_{\mu\tau\omega\nu},   \quad    \widetilde{{\mathcal
R}}=\frac{1}{2}\epsilon^{\nu\mu\tau\omega}{\mathcal
R}_{\mu\tau\omega\nu}.
\end{equation}

Under the condition that the variation of the dual action
vanishes, {\it i.e.},
$\delta\int_{\Omega}\sqrt{-g}\widetilde{{\mathcal R}}{\rm
d}\Omega=0$, one can arrive at an antisymmetric dual Einstein
tensor
\begin{equation}
\widetilde{{\mathcal G}}_{\mu\nu}=\widetilde{{\mathcal
R}}_{\mu\nu}-\widetilde{{\mathcal R}}_{\nu\mu},
\end{equation}
where the dual Ricci tensor $\widetilde{{\mathcal
R}}_{\mu\nu}=g^{\tau\omega}\widetilde{{\mathcal
R}}_{\mu\tau\omega\nu}$. In addition, according to the definition
of the dual Riemannian curvature tensor (\ref{dualRiemann}),
$\widetilde{{\mathcal R}}_{\mu\nu}$ can be rewritten in the form
\begin{equation}
\widetilde{{\mathcal R}}_{\mu\nu}=\frac{1}{2}\epsilon_{\mu}^{\ \
\lambda\sigma\omega}{\mathcal R}_{\lambda\sigma\omega\nu}.
\end{equation}

It should be noted that the NUT metric (with a vanishing
Schwarzschild mass parameter) is in fact the solution to the
source-free Einstein's gravitational field equation (vacuum
equation). Einstein's source-free field equation can be viewed as
a Bianchi identity seen from the point of view of the theoretical
framework of dual curvature tensors \cite{Shen2}. In other words,
the gravitomagnetic charge is required of a gravitational field
equation of its own for treating the dynamics in the spacetime in
which the gravitomagnetic monopole is present. The possible field
equation may be of the form
\begin{equation}
\widetilde{{\mathcal G}}_{\mu\nu}=S_{\mu\nu},
\label{fieldequation}
\end{equation}
where $S_{\mu\nu}$ is an antisymmetric source tensor of the
gravitomagnetic matter \cite{Shen,Shen2}, which will be discussed
in detail below.

The physical meanings of the field equation (\ref{fieldequation})
in the weak-field approximation is explicit: if the
gravitomagnetic vector potentials, $g_{0\lambda}$ and $g_{\sigma
0}$ ($\lambda, \sigma=0-3$), can be rewritten as $a_{\lambda}$ and
$a_{\sigma}$, respectively, then by using the formula
$\widetilde{{\mathcal R}}_{\mu\nu}=(1/4)\epsilon_{\mu}^{\ \
\lambda\sigma\omega}\partial_{\omega}\left(\partial_{\sigma}g_{\nu\lambda}-\partial_{\lambda}g_{\sigma\nu}\right)$,
one can show that the $({\mu 0})$ component of the dual Ricci
tensor in the weak-field approximation takes the form
$\widetilde{{\mathcal R}}_{\mu 0}=(1/4)\epsilon_{\mu}^{\ \
\lambda\sigma\omega}\partial_{\omega}f_{\sigma\lambda}$, where the
weakly gravitational field ``tensor'' is of the form
\begin{equation}
f_{\sigma\lambda}=\partial_{\sigma}a_{\lambda}-\partial_{\lambda}a_{\sigma}.
\label{fieldtensor}
\end{equation}
Apparently, in the weak-field approximation the quantity
$\widetilde{{\mathcal R}}_{\mu 0}$ is exactly analogous to the
divergence of the dual electromagnetic field tensor,
$\partial_{\omega}\widetilde{\mathcal F}_{\mu}^{\ \ \omega}$ with
$\widetilde{\mathcal F}_{\mu}^{\ \ \omega}=(1/2)\epsilon_{\mu}^{\
\ \lambda\sigma\omega}{\mathcal F}_{\sigma\lambda}$. Such a
divergence term appears on the left-handed side of the
electromagnetic field equation of magnetic charge
\cite{Lynden-Bell,Zeleny}. In a word, as the left-handed side of
Eq. (\ref{fieldequation}) contains a divergence term of dual field
``tensor'', Eq. (\ref{fieldequation}) may indeed be viewed as a
gravitational field equation of gravitomagnetic matter of its own.

As a kind of topological charge in spacetime, gravitomagnetic
charge will unavoidably result in a nonanalyticity of the metric
functions. For this reason, one can obtain the following two
asymmetric Ricci tensors
\begin{equation}
{\Re}_{\mu\nu}=g^{\alpha\beta}{\mathcal R}_{\alpha\mu\nu\beta},
\qquad {\Re}'_{\mu\nu}=g^{\alpha\beta}{\mathcal
R}_{\mu\alpha\beta\nu}.
\end{equation}
The connection between these two Ricci tensors is given by
\begin{equation}
{\Re}'_{\mu\nu}={\Re}_{\mu\nu}-g^{\alpha\beta}
\left(\partial_{\nu}\partial_{\beta}-\partial_{\beta}\partial_{\nu}\right)g_{\alpha\mu},
\end{equation}
where the expression
$\left(\partial_{\nu}\partial_{\beta}-\partial_{\beta}\partial_{\nu}\right)g_{\alpha\mu}$
no longer vanishes if the metric is {\it nonanalytic}. In spite of
this, the contractions of these two asymmetric Ricci tensors
yields the same curvature scalar, {\it i.e.}, ${\mathcal
R}=g^{\mu\nu}{\Re}_{\mu\nu}=g^{\mu\nu}{\Re}'_{\mu\nu}$. It follows
that once the gravitomagnetic monopole is absent and the metric
tensor is then analytic, the two Ricci tensors are the same
(${\Re}_{\mu\nu}={\Re}'_{\mu\nu}$), and consequently the symmetric
property of ${\Re}_{\mu\nu}$ in indices (${\mu, \nu}$) will be
recovered.

How can we construct a symmetric Ricci tensor within the framework
of {\it nonanalytic} metric? Note that the Ricci tensor is in
general the contraction from the curvature tensor ${\mathcal
R}_{\alpha\mu\nu\beta}$ that is antisymmetric in both ($\alpha,
\mu$) and $(\nu, \beta)$. As stated above, there are two
candidates ${\Re}_{\mu\nu}, {\Re}'_{\mu\nu}$. We can construct a
fourth-rank tensor ${\Re}_{\alpha\mu\nu\beta}=(1/2)\left({\mathcal
R}_{\alpha\mu\nu\beta}+{\mathcal R}_{\mu\alpha\beta\nu}\right)$.
It can be readily verified that this tensor is truly antisymmetric
in indices ($\alpha, \mu$) and $(\nu, \beta)$ seen from the
following relations:
${\Re}_{\mu\alpha\nu\beta}=(1/2)\left({\mathcal
R}_{\mu\alpha\nu\beta}+{\mathcal
R}_{\alpha\mu\beta\nu}\right)=-(1/2)\left({\mathcal
R}_{\mu\alpha\beta\nu}+{\mathcal
R}_{\alpha\mu\nu\beta}\right)=-{\Re}_{\alpha\mu\nu\beta}$. Thus
the contraction $g^{\alpha\beta}{\Re}_{\alpha\mu\nu\beta}$ may be
considered a Ricci tensor. However, this tensor is not symmetric
in indices ($\mu, \nu$). In order to obtain a symmetric Ricci
tensor, we should symmetrize it, namely, the following form
\begin{equation}
{\mathcal
R}_{\mu\nu}=\frac{1}{4}\left[\left({\Re}_{\mu\nu}+{\Re}'_{\mu\nu}\right)
+\left({\Re}_{\nu\mu}+{\Re}'_{\nu\mu}\right)\right]
\label{Riccicandidadte}
\end{equation}
is chosen as the required symmetric Ricci tensor. If the partial
derivatives can commute, expression (\ref{Riccicandidadte}) will
be reduced to the regular form in Riemannian geometry, the metric
of which is analytic.

Now let us return to the field equation (\ref{fieldequation}). The
problem left to us is to construct the so-called source tensor
$S_{\mu\nu}$. Note that the dual tensor $\widetilde{{\mathcal
G}}_{\mu\nu}$ is antisymmetric in indices (${\mu, \nu}$), so is
the source tensor. As a tentative consideration, here we only
analyze the spinor field $\psi$ that characterizes the {\it spinor
gravitomagnetic matter}. Clearly, the following two tensors which
possess an antisymmetric property can be immediately achieved: one
is ${\mathcal
A}_{\mu\nu}=\bar{\psi}\left(\gamma_{\mu}\gamma_{\nu}-\gamma_{\nu}\gamma_{\mu}\right)\psi$,
and the other is its dual, $(1/2)\epsilon_{\mu\nu}^{\ \ \
\theta\tau}{\mathcal A}_{\theta\tau}$. Here the Dirac matrices
$\gamma_{\mu}$'s are defined through the relation
$\gamma_{\mu}=e_{\mu}^{\ a}\gamma_{a}$, where $e_{\mu}^{\ a}$
denotes the vierbein field and the Latin suffixes $a$'s stand for
the indices of the coordinates in the flat Minkowski spacetime. It
is well known that the spin (or the spinning gravitomagnetic
moment) of a massive particle is the source of a gravitomagnetic
field. Likewise, the spin of a gravitomagnetic monopole (with a
{\it dual} mass) may be the source of a gravitoelectric field
(Newtonian gravitational field). If, for example, the quantity
${\mathcal A}_{\mu\nu}$ acts as a source tensor, the spinning
gravitomagnetic moment of dual mass will produce {\it
gravitoelectric} fields. The dual Einstein tensor
$\widetilde{{\mathcal G}}_{\mu\nu}$, which, as stated above,
contains a divergence term of the dual weakly-gravitational field
``tensor'', is, however, of the ``magnetic'' type. Thus such a
source tensor should be removed for this reason. Since ${\mathcal
A}_{\mu\nu}$ can produce gravitoelectric fields, its dual may in
turn produce gravitomagnetic fields instead. Hence, a reasonable
choice may be the adoption of the dual, $(1/2)\epsilon_{\mu\nu}^{\
\ \ \theta\tau}{\mathcal A}_{\theta\tau}$, as the source tensor of
Eq. (\ref{fieldequation}).

In view of the above discussion, the simplest form of the
gravitational field equation of gravitomagnetic charge ({\it
e.g.}, the spinor dual matter) reads
\begin{equation}
\widetilde{{\mathcal G}}_{\mu\nu}=\kappa \epsilon_{\mu\nu}^{\ \ \
\theta\tau}\bar{\psi}\gamma_{\theta}\gamma_{\tau}\psi,
\label{fieldequation2}
\end{equation}
where $\kappa$ is a certain coupling coefficient.

To see the connection between matter and dual matter in
gravitational field equations, we multiply the two sides of Eq.
(\ref{fieldequation2}) by a fully antisymmetric Levi-Civita tensor
$\epsilon^{\mu\nu\alpha\beta}$, and then arrive at
\begin{equation}
\frac{1}{2}\epsilon^{\mu\nu\alpha\beta}\widetilde{{\mathcal
G}}_{\mu\nu}=-\kappa\bar{\psi}\left(\gamma^{\alpha}\gamma^{\beta}-\gamma^{\beta}\gamma^{\alpha}\right)\psi.
\label{fieldequation3}
\end{equation}
Taking account of the relation  \cite{Shen2}
\begin{equation}
\frac{1}{2}\epsilon^{\mu\nu\alpha\beta}\widetilde{{\mathcal
G}}_{\mu\nu}=-\frac{1}{2}\left[\left({\Re}^{\alpha\beta}+{\Re}'^{\alpha\beta}\right)
-\left({\Re}^{\beta\alpha}+{\Re}'^{\beta\alpha}\right)\right],
\end{equation}
one can rewrite Eq. (\ref{fieldequation3}) as a set of equations
\begin{eqnarray}
& &
  \frac{1}{2}\left({\Re}^{\alpha\beta}+{\Re}'^{\alpha\beta}\right)
=\kappa\bar{\psi}\gamma^{\alpha}\gamma^{\beta}\psi+s^{\alpha\beta},
\nonumber  \\
& &
 \frac{1}{2}\left({\Re}^{\beta\alpha}+{\Re}'^{\beta\alpha}\right)
=\kappa\bar{\psi}\gamma^{\beta}\gamma^{\alpha}\psi+s^{\beta\alpha},
\label{fieldequation4}
\end{eqnarray}
where $s^{\alpha\beta}$ is a certain symmetric tensor, the
physical meanings of which will be revealed in the following. If
we take into account the relations (\ref{Riccicandidadte}) and
$\gamma^{\alpha}\gamma^{\beta}+\gamma^{\beta}\gamma^{\alpha}=2g^{\alpha\beta}$,
we can obtain a formula
\begin{equation}
{\mathcal R}^{\alpha\beta}=\kappa\bar{\psi}\psi
g^{\alpha\beta}+s^{\alpha\beta}            \label{fieldequation5}
\end{equation}
from Eqs. (\ref{fieldequation4}). According to Einstein's field
equation, the symmetric tensor $s^{\alpha\beta}$ may act as the
source tensor of gravitoelectric matter and thus takes the form
$\Lambda g^{\alpha\beta}-(8\pi
G/c^{4})\left(T^{\alpha\beta}-g^{\alpha\beta}T/2\right)$, where
$\Lambda$, $G$ and $c$ denote the cosmological constant, Newtonian
gravitational constant and speed of light in vacuum, respectively.
$T^{\alpha\beta}$ is the energy-momentum tensor of matter
(gravitoelectric matter). The connection between gravitomagnetic
and gravitoelectric matter in the gravitational field equations is
thus demonstrated by comparing Eq. (\ref{fieldequation2}) with Eq.
(\ref{fieldequation5}).

For simplicity, the gravitoelectric matter is here assumed to be
absent, and in consequence the energy-momentum tensor
$T^{\alpha\beta}$ is taken to be zero (and hence $T=0$). By
introducing a parameter $\lambda$, Eq. (\ref{fieldequation5}) can
be rewritten as
\begin{equation}
{\mathcal R}^{\alpha\beta}=\left(\Lambda+\lambda\right)
g^{\alpha\beta}      \label{Einsteinequation}
\end{equation}
with $\lambda=\kappa\bar{\psi}\psi$. It follows that Eq.
(\ref{Einsteinequation}) is in exact analogy with Einstein's
vacuum field equation with a modified cosmological constant
$\Lambda+\lambda$. Hence, the term $\kappa\bar{\psi}\psi$
associated with the gravitomagnetic matter can play a role
analogous to the cosmological constant, and the gravitomagnetic
charge may be viewed as one of the potential origins of the
cosmological constant (or dark energy \cite{darkenergy}).

The cosmological constant problem is that why the observed value
of the vacuum energy density is so small: the theoretical value of
the cosmological constant resulting from the quantum vacuum
fluctuation is
\begin{equation}
\Lambda_{\rm th}\simeq \frac{c^{3}}{\pi G\hbar},
\label{theo}
\end{equation}
where $\hbar$ denotes the Planck constant. However, the ratio of
experimental value to the theoretical one is only $10^{-120}$
\cite{Weinberg2}. During the past 40 years, in an attempt to deal
with the cosmological constant problem, many theoretical
mechanisms \cite{Weinberg2} such as the adjustment mechanism
\cite{Kim,Dine}, changing gravity \cite{Dragon}, quantum cosmology
and viewpoints of supersymmetry \cite{Hawking}, supergravity as
well as superstrings \cite{Witten} were proposed. More recently,
since some astrophysical observations ({\it e.g.}, Type Ia
supernova observations  \cite{Nature}) showed that the large scale
mean pressure of our present universe is negative suggesting a
positive but small cosmological constant, and that the universe is
therefore presently undergoing an accelerating expansion
\cite{Nature}, a large number of theories and viewpoints have been
put forward to resolve the cosmological constant problem. These
include the back reaction of cosmological perturbations
\cite{back}, QCD trace anomaly \cite{Sch}, contribution of
Kaluza-Klein modes to vacuum energy \cite{contribution},
five-dimensional unification of cosmological constant and photon
mass \cite{five}, nonlocal quantum gravity \cite{Moffat}, quantum
microstructure of spacetime \cite{Padmanabhan}, relaxation of the
cosmological constant in a movable brane world \cite{relaxation},
effects of minimal length uncertainty relation (using the modified
commutation relation $[q, p]$) on the density of states
\cite{Chang} as well as comoving suppression mechanism
\cite{comvingshen}. Although none of these theories achieves a
definite and reliable success in dealing with the problem, they
drew inspiration from the current knowledge of physics and
suggested a variety of possibilities to overcome this difficulty,
which enlighten us on this subject. As a new possibility, here we
also suggest a new scheme based on the concept of gravitomagnetic
monopole to treat this problem.

First we assume that there exists a certain creation mechanism of
gravitomagnetic charge in the gauge theory (for example, the
interaction related to the Chern-Simons gauge fields
\cite{Hagen,Duan}). Then as demonstrated in Eq.
(\ref{Einsteinequation}), if the vacuum fluctuation of the
gravitomagnetic matter can truly contributes to the cosmological
constant, the theoretical value of the cosmological constant may
be dramatically suppressed by a large number of orders of
magnitude, provided that the parameter $\lambda$ in Eq.
(\ref{Einsteinequation}) has a minus sign. The present
interpretation for the smallness of cosmological constant is
suggested in principle but not in detail, since the coupling
coefficient $\kappa$ in Eqs. (\ref{fieldequation5}) and
(\ref{Einsteinequation}) cannot be determined by the mechanism
itself. But, there are some clues, which can help us to extract
some information on the coupling coefficient $\kappa$. It follows
from Eq. (\ref{fieldequation5}) that the dimension of $\kappa$ is
$[L]$, since the dimensions of $\psi$ ($\bar{\psi}$) and
${\mathcal R}^{\alpha\beta}$ are $[L^{-\frac{3}{2}}]$ and
$[L^{-2}]$, respectively. If $\kappa$ can be constructed in terms
of the fundamental physical constants such as $G$, $\hbar$ and
$c$, then the only expression that has a dimension of $[L]$ is
$\sqrt{G\hbar/c^{3}}$. Therefore, we assume that $\kappa\simeq
-\sqrt{G\hbar/c^{3}}$ (the minus sign may result from the
Meissner-like effect discussed below). Generally speaking,
$\bar{\psi}\psi$ can be regarded as the phase space density of
vacuum fluctuation fields. By using the formula
$\bar{\psi}\psi\sim [2/\left(2\pi \hbar\right)^{3}]\int^{{p}_{\rm
P}/\hbar}{\rm d}^{3}{\bf k}$, where ${p}_{\rm P}$ denotes the
Plank momentum, one can obtain
\begin{equation}
\bar{\psi}\psi\simeq
\frac{c^{\frac{9}{2}}}{3\pi^{2}(G\hbar)^{\frac{3}{2}}}.
\label{rho}
\end{equation}
Thus $\lambda=\kappa\bar{\psi}\psi\simeq -c^{3}/(3\pi^{2}
G\hbar)$, the modulus of which is compared to the previous
theoretical value of the cosmological constant in expression
(\ref{theo}). This, therefore, means that there is a possibility
for $\lambda$ to dramatically modify the value of the cosmological
constant, if the dual matter is truly present and its contribution
is then taken into account.

On the other hand, if the vacuum fluctuation field can be thought
of as a perfect fluid, we can find a Meissner-type mechanism from
the dynamical point of view: specifically, the gravitoelectric
field (Newtonian gravitational field) produced by the mass of the
quantum vacuum fluctuation can be exactly cancelled by the
gravitoelectric field resulting from the induced dual mass current
at vacuum-fluctuation level, and in turn, the gravitomagnetic
field produced by the gravitomagnetic charge (dual mass) of the
vacuum quantum fluctuation can also be cancelled exactly by the
gravitomagnetic field resulting from the induced mass current at
vacuum-fluctuation level. Thus, the gravitational effect of the
cosmological constant $\Lambda$ may, in principle, be eliminated
(by many orders of magnitude) by the contribution, $\lambda$, of
the gravitomagnetic charge (dual mass).

Finally, let us return to the topic of nonanalytic metric produced
by the gravitomagnetic monopole. We note that there are
nonanalytic electromagnetic vector potentials in the case of Dirac
monopole \cite{Dirac}. Wu and Yang found that only the phase
factor (rather than field strengths or electromagnetic potentials)
is more physically meaningful and therefore constitutes an
intrinsic and complete description of electromagnetism \cite{Wu}.
As for the case of Dirac monopole, the nonintegrable phase factor
can be viewed as a fundamental concept to describe the
electromagnetism of magnetic monopole. However, the nonintegrable
phase factor with nonanalytic vector potentials becomes undefined
if the path goes through a singularity. In order to resolve this
difficulty, Wu and Yang used a fibre-bundle construction to avoid
the singularity of the nonanalytic vector potentials \cite{Wu}. A
question whether a fibre bundle reformulation of the theory of
gravitomagnetic monopoles, which can avoid the singularity of
nonanalytic metric, exists is therefore left to us. As stated
above, if the gravitomagnetic monopole is at rest, the $(\mu 0)$
component of the dual Ricci tensor takes the form
$\widetilde{{\mathcal R}}_{\mu 0}=(1/4)\epsilon_{\mu}^{\ \
\lambda\sigma\omega}\partial_{\omega}f_{\sigma\lambda}$, where the
weakly gravitational field strength ``tensor'' is defined by Eq.
(\ref{fieldtensor}). This, therefore, means that the $(\mu
0)$-component dual Ricci tensor has exactly a same mathematical
structure as the divergence of dual electromagnetic field tensor.
Thus, at least for the case of gravitomagnetic monopole at rest,
where the static gravitomagnetic field is produced, there exists a
similar fibre-bundle construction to define the singularity-free
regions and to avoid the corresponding Dirac strings that arise
from gravitomagnetic monopoles. The problem regarding whether the
singularity-free regions can be defined or not for the other
components of the dual Ricci tensor $\widetilde{{\mathcal R}}_{\mu
\nu}$ with $\nu\neq 0$ deserves further consideration.

To summarize, since less attention was paid to the nonanalyticity
of metric in the literature, we presented a connection between
curvature tensors and dual curvature tensors in terms of the
nonanalytic metric, and then discussed the relationship between
matter and dual matter in gravitational field equations. A new
possibility, which may account for the smallness of the
cosmological constant, was proposed by using the
gravitoelectromagnetic cancellation mechanism resulting from the
dual matter at vacuum-fluctuation level. Such a dual matter is
viewed as a perfect fluid. Although at present there are no
evidences for the existence of such a topological dual mass, its
dynamics and related topics still deserve further investigation
theoretically. This may enables us to better understand
topological (global) phenomena and to find more such phenomena in
gravity theory.


\begin{references}
\bibitem{Overhauser} A.W. Overhauser, R. Colella, Phys. Rev. Lett. 33 (1974) 1237.

\bibitem{Werner}  S.A. Werner, J.L. Staudenmann, R. Colella, Phys.
Rev. Lett 42 (1979) 1103.

\bibitem{Anandan} J. Anandan, Phys. Rev. D 15 (1977) 1448.

\bibitem{Dresden} M. Dresden, C.N. Yang, Phys. Rev. D 20 (1979) 1846.

\bibitem{Furtado} C. Furtado, V.B. Bezerra, Phys. Rev. D 62 (2000) 045003.

\bibitem{Wagh} A.G. Wagh, G. Badurek, V.C. Rakhecha, et al., Phys. Lett. A 268 (2000) 209.

\bibitem{Shenprb} J.Q. Shen, S.L. He, Phys. Rev. B 68 (2003) 195421.


\bibitem{Hehl}  F.W. Hehl, W.T. Ni, Phys. Rev. D 42 (1990) 2045.

\bibitem{Ahmedov}  B.J. Ahmedov, Phys. Lett. A 256 (1999) 9.


\bibitem{Mashhoon1}  B. Mashhoon, Phys. Lett. A 173 (1993) 347.

\bibitem{Mashhoon2}  B. Mashhoon, Gen. Relativ. Gravit. 31 (1999) 681.

\bibitem{Mashhoon3}  B. Mashhoon, Class. Quant. Gravit. 17 (2000) 2399.

\bibitem{Shenscript}  J.Q. Shen, H.Y. Zhu, S.L. Shi, J. Li, Phys. Scr. 65 (2002) 465.

\bibitem{Shen3}  J.Q. Shen, Phys. Rev. D 70 (2004) 067501.


\bibitem{Newman} E.T. Newman, L. Tamburino, T. Unti, J. Math. Phys. 4 (1963) 915.

\bibitem{Demiansky} M. Demianski, E.T. Newman, Bull. Acad. Pol. Sci. Ser. Math. Astron. Phys. XIV (1966) 653.


\bibitem{Nouri-Zonoz} M. Nouri-Zonoz, xxx.lanl.gov (1997) gr-qc/9706015.

\bibitem{Zimmerman} R.L. Zimmerman, B.Y. Shahir, Gen. Relativ. Gravit.
21 (1989) 821.

\bibitem{Lynden-Bell}  D. Lynden-Bell, M. Nouri-Zonoz, Rev. Mod. Phys. 70 (1998) 427.

\bibitem{Miller} J.G. Miller, J. Math. Phys. 14 (1973) 486.

\bibitem{Bini2002}  D. Bini, Class. Quant. Gravit. 19 (2002) 5265.

\bibitem{Bini1} D. Bini, C. Cherubini, R.T. Jantzen,
B. Mashhoon, Phys. Rev. D 67 (2003) 084013.


\bibitem{Bini2}  D. Bini, C. Cherubini, R.T. Jantzen, Class. Quant. Gravit. 19 (2002) 5481.


\bibitem{Dowker} J.S. Dowker, J.A. Roche, Proc. Phys. Soc. London 92 (1967) 1.

\bibitem{Dowker2}  J.S. Dowker, Gen. Relativ. Gravit. 5 (1974) 603.

\bibitem{Zee}  A. Zee, Phys. Rev. Lett. 55 (1985) 2379.

\bibitem{Shen} J.Q. Shen, Gen. Relativ. Gravit. 34 (2002) 1423.


\bibitem{Shen2} J.Q. Shen, Ann. Phys. (Leipzig) 13 (2004) 532.

\bibitem{Weinberg2}    S. Weinberg, Rev. Mod. Phys. 61 (1988) 1.

\bibitem{Zeleny} W.B. Zeleny, Am. J. Phys. 59 (1991) 412.

\bibitem{darkenergy} S. Perlmutter, M. Turner, M. White, Phys. Rev. Lett. 83 (1999) 670.

\bibitem{Kim}  J. Kim, Phys. Rev. Lett. 43 (1979) 103.

\bibitem{Dine} M. Dine, W. Fischler, M. Srednicki, Phys. Lett. B 104 (1981) 199.

\bibitem{Dragon} W. Buchm\"{u}ller, N. Dragon, Phys. Lett. B 207 (1988) 292.

\bibitem{Hawking} S.W. Hawking, Phys. Lett. B 134 (1984) 403.

\bibitem{Witten} E. Witten, Lecture at DM 2000, Marina del Rey [xxx.lanl.gov (2000)
hep-ph/0002297].


\bibitem{Nature}  S. Perlmutter, G. Aldering, M.D. Valle, et al., Nature 391 (1998) 51;
S. Perlmutter, G. Aldering, G. Goldhaber, et al., Astrophys. J.
517 (1999) 565; A.G. Riess, A.V. Filippenko, P. Challis, et al.,
Astrophys. J. 116 (1998) 1009.

\bibitem{back} R.H. Brandenberger, xxx.lanl.gov(2002) hep-th/0210165.

\bibitem{Sch}  R. Sch\"{u}tzhold, xxx.lanl.gov (2002) gr-qc/0204018.

\bibitem{contribution} A. Gupta, xxx.lanl.gov (2002) hep-th/0210069.

\bibitem{five} C. Kohler, xxx.lanl.gov (2002) gr-qc/0202076.

\bibitem{Moffat} J.W. Moffat, AIP Conf. Proc. 646 (2003) 130.

\bibitem{Padmanabhan} T. Padmanabhan, Class. Quan. Gravit. 19 (2002)
L167.

\bibitem{relaxation} S. Khlebnikov, xxx.lanl.gov (2002) hep-th/0207258.

\bibitem{Chang} L.N. Chang, D. Minic, N. Okamura, T. Takeuchi, xxx.lanl.gov (2002) hep-th/0201017.

\bibitem{comvingshen} J.Q. Shen, xxx.lanl.gov (2004)
gr-qc/0401077.


\bibitem{Hagen} C.R. Hagen, Phys. Rev. Lett. 68 (1992) 3821.

\bibitem{Duan} Y.S. Duan, X. Liu, L.B. Fu, Commun. Theor. Phys. (Beijing,
China) 40 (2003) 447.

\bibitem{Dirac} P.A.M. Dirac, Proc. R. Soc. London A 133 (1931)
60.

\bibitem{Wu} T.T. Wu, C.N. Yang, Phys. Rev. D 12 (1975) 3845.
\end{references}
\end{document}